\documentclass[twocolumn,showpacs,amsmath,amssymb,aps]{revtex4}
\usepackage{mathrsfs}
\usepackage[dvips]{graphicx}

\begin{document}

\title{Vacuum Fluctuations and the Cosmological Constant}
\author{Shi Qi}
\email{qishi@chenwang.nju.edu.cn}
\affiliation{Department of Physics, Nanjing University, Nanjing 210093, China \\
  and Joint Center for Particle, Nuclear Physics and Cosmology, Nanjing University, Nanjing 210093, China
}
\date{\today}
\begin{abstract}
The hypothesis is proposed that under the approximation that the quantum
equations of motion reduce to the classical ones,
the quantum vacuum also reduces to the classical vacuum---the empty
space. The vacuum energy of QED is studied under
this hypothesis. A possible solution to the cosmological constant
problem is provided and a kind of parameterization of the cosmological
``constant'' is derived.
\end{abstract}

\pacs{11.10.-z, 98.80.Cq}

\maketitle

\section{Introduction}

If the physics of the vacuum looks the same to any inertial
observer, as it should be, its contribution to the energy-momentum tensor must be invariant under
Lorentz transformation. The only class of tensors that have such a property is
a multiple of the metric tensor. So the contribution of the vacuum takes the same
form as the cosmological constant term in the Einstein field equation. 
This result can also be derived in another way.
we know the proper volume element of space with the Robertson-Walker metric
\begin{equation}
  \label{eq:RWmetric}
  \mathrm{d}s^2=\mathrm{d}t^2-a^2(t)
  \left(  \frac{\mathrm{d}r^2}{1-kr^2}+r^2\mathrm{d}\theta^2+r^2\sin^2\theta\mathrm{d}\phi^2 \right)
\end{equation}
is given by
\begin{equation}
  \label{eq:proper_volume_element}
  \mathrm{d}V = a^3(t) \frac{r^2 \sin \theta}{\sqrt{1-k r^2}}
  \mathrm{d} r \mathrm{d}\theta \mathrm{d} \phi
\end{equation}
The proper volume evolves as the scale factor $a(t)$ evolves.
Increasing $a(t)$ corresponds to creation of space. If one
treats the vacuum as an ideal fluid, then in the momentarily comoving
reference frame(MCRF) its energy-momentum tensor takes the form
diag$(\rho_{vac}, p_{vac}, p_{vac}, p_{vac})$.
Unlike other ideal fluids, the vacuum associates with the space.
It is produced immediately after creation of space.
Filling the created space of proper volume $\delta V$ with the vacuum
would cause an increase of energy of $\rho_{vac}\delta V$.
Energy conservation requires $-p_{vac}\delta V = \rho_{vac}\delta V$,
or $p_{vac}=-\rho_{vac}$. So the energy-momentum tensor of the vacuum
is $\rho_{vac}$diag$(1, -1, -1, -1)$, a multiple of the metric tensor.
The vacuum fluctuations themselves are real. However,
a straightforward calculation of the energy density of the vacuum leads to
a infinitely large cosmological constant. A cutoff at reasonable momentum
yields a vacuum energy density that is still many orders of magnitude
larger than observation. This is the well-known
cosmological constant problem,
as reviewed by many authors~\cite{zeldovich1968}~\cite{zeldovich1981}
~\cite{Weinberg:1988cp}~\cite{Carroll:1991mt}~\cite{Sahni:1999gb}
~\cite{Carroll:2000fy}~\cite{Rugh:2000ji}~\cite{Peebles:2002gy}
~\cite{Padmanabhan:2002ji}~\cite{Padmanabhan:2004av}~\cite{Nobbenhuis:2004wn}.

With the development of supersymmetry,
Zumino~\cite{Zumino:1974bg} pointed out that unbroken supersymmetry implies
a vanishing vacuum energy density. But the trouble is that supersymmetry
is broken at low temperatures such as in today's universe.
See \cite{Weinberg:1988cp} for a detailed discussion of this aspect of the problem.

Due to the huge discrepancy between theory and observation, a lot of possible
solutions have been proposed. See~\cite{Weinberg:1988cp}~\cite{Carroll:1991mt}
~\cite{Carroll:2000fy}~\cite{Rugh:2000ji}~\cite{Padmanabhan:2002ji}
~\cite{Padmanabhan:2004av}~\cite{Nobbenhuis:2004wn} for reviews.
See also~\cite{Shen:2005nr} for another possible cancellation mechanism.

In this letter, we present another possible
way of thinking about the problem.

\section{The basic hypothesis}

Consider a theory of a real scalar field with Lagrangian density
\begin{equation}
  \label{eq:Lagrangian}
  \mathscr{L}=\frac{1}{2} \partial_{\mu} \phi \partial^{\mu} \phi
  - \frac{1}{2}m^2\phi^2 - \frac{1}{4!}g\phi^4
\end{equation}
The energy-momentum density of the vacuum
\begin{equation}
  \label{eq:em}
  T_{vac}^{\mu\nu}= \langle 0 \vert T^{\mu\nu} \vert 0 \rangle
\end{equation}
where $T^{\mu\nu}=\partial^{\mu}\phi \partial^{\nu}\phi - g^{\mu\nu}\mathscr{L}$,
can be written in terms of Green functions by making use of
\begin{eqnarray}
  \label{eq:phi4-quadratic} 
  \langle 0 \vert \phi^2 \vert 0 \rangle & = & \lim_{x-y \to 0} \langle 0 \vert T\{ \phi(x)\phi(y) \} \vert 0 \rangle 
  \\
  %%%%%%%%%%%%%%%%%%%%%%%%%%%%%%%%%%%%%%%%%%%%%%%%%%
  \label{eq:phi4-partial} 
  \langle 0 \vert \partial^{\mu}\phi \partial^{\nu}\phi \vert 0 \rangle
  & = & \lim_{x-y \to 0}\frac{\partial}{\partial x_{\mu}} \frac{\partial}{\partial y_{\nu}}
  \langle 0 \vert T\{ \phi(x)\phi(y) \} \vert 0 \rangle
  \\
  %%%%%%%%%%%%%%%%%%%%%%%%%%%%%%%%%%%%%%%%%%%%%%%%%%%
  \label{eq:phi4-quartic}
  \langle 0 \vert \phi^4 \vert 0 \rangle & = & \lim_{x_1, x_2, x_3, x_4 \to x}
  \nonumber \\
  & & \langle 0 \vert T\{ \phi(x_1)\phi(x_2)\phi(x_3)\phi(x_4) \} \vert 0 \rangle 
\end{eqnarray}
(Eq.~(\ref{eq:phi4-quadratic}) and Eq.~(\ref{eq:phi4-quartic}) are self-evident.
The proof of Eq.~(\ref{eq:phi4-partial}) is put in the appendix)
For example, for the theory of free real scalar field ($g=0$ in Eq.~(\ref{eq:Lagrangian}) )
\begin{equation}
  \langle 0 \vert T\{ \phi(x)\phi(y) \} \vert 0 \rangle = \int \frac{\mathrm{d}^4 k}{(2\pi)^4}
  \frac{i}{k^2-m^2+i\epsilon} e^{-ik \cdot (x-y)}
\end{equation}
the vacuum energy-momentum density is
\begin{eqnarray}
  T_{vac}^{\mu\nu} & = & \lim_{x-y \to 0} \int \frac{\mathrm{d}^4 k}{(2\pi)^4}  \frac{i}{k^2-m^2+i\epsilon} 
  \nonumber \\
  &  &	\times [k^{\mu}k^{\nu}-\frac{1}{2}g^{\mu\nu}(k^2-m^2)] e^{-ik \cdot (x-y)} \nonumber \\
  %%%%%%%%%%%%%%%%%%%%%%%%%%
  & = & \int \frac{\mathrm{d}^3 k}{(2\pi)^3} \frac{1}{2\omega_{\vec{k}}}
  [k^{\mu}k^{\nu}    \nonumber \\
  &  & \left. -\frac{1}{2}g^{\mu\nu}(k^2-m^2)] \right\vert_{k^0=\pm \omega_{\vec{k}}=\pm\sqrt{\vec{k}^2+m^2}}
  \nonumber \\
  & = & \left. \int \frac{\mathrm{d}^3 k}{(2\pi)^3} \frac{1}{2\omega_{\vec{k}}}
  k^{\mu}k^{\nu} \right\vert_{k^0=\pm \omega_{\vec{k}}=\pm\sqrt{\vec{k}^2+m^2}}
\end{eqnarray}
(Here we have assumed there is a well behaved cutoff function in the integrand,
which is not written out explicitly, so that we can carry out the
contour integration over $k^{0}$)
which is the same result as just to substitute
\begin{equation}
  \phi(x)=\int \frac{\mathrm{d}^3 k}{(2\pi)^3} \frac{1}{2\omega_{\vec{k}}}
  [a(\vec{k})e^{-ik \cdot x} + a^{\dagger}(\vec{k})e^{ik \cdot x}]
\end{equation}
into Eq.~(\ref{eq:em}).
In particular, 
\begin{equation}
\label{eq:ZeroPointEnergy}
  T_{vac}^{00} = \int \frac{\mathrm{d}^3 k}{(2\pi)^3} \frac{\omega_{\vec{k}}}{2}
\end{equation}
is the zero point energy density.
In particle physics, what matters is the relative energy, not the absolute
value, and such infinities are eliminated by normal ordering.
However, in general relativity, energy-momentum tensor is the source of
space-time curvature and vacuum energy-momentum should be taken into account.
Unfortunately, Eq.~(\ref{eq:ZeroPointEnergy}) is divergent and any reasonable
cutoff leads to a cosmological constant which is unacceptably high.

Recall that if only tree level Feynman graphs are considered,
the quantum effective Lagrangian is the same as
the classical Lagrangian and quantum field equations as classical
field equations. We put it forward as a hypothesis that under the
approximation that the quantum equations of motion reduce to the
classical ones, the quantum vacuum also reduces to the
classical one. In classical theories, the vacuum is just the empty
space and there is no vacuum fluctuations. So for the case of vacuum
energy-momentum, it is equivalent to assuming that the part of vacuum
energy-momentum that arises from tree level Green functions is
unobservable in principle, and therefore does not contribute to gravitation.
If we view the $\Lambda$ term as a quantum correction of matter fields
to Einstein field equation of general relativity without $\Lambda$ term, then 
it follows from the above hypothesis that such a $\Lambda$-term quantum correction
vanishes for tree level Green functions, which is analogous to the fact that
quantum corrections to classical theories arise from loop Feynman graphs
in quantum field theory. 

After excluding tree level Green functions, the lowest
order contribution in $g$ (quadratic in $g$) of vacuum energy-momentum density arises only from 
$\partial^{\mu}\phi \partial^{\nu}\phi - g^{\mu\nu}
(\frac{1}{2} \partial_{\mu} \phi \partial^{\mu} \phi - \frac{1}{2}m^2\phi^2)$
in $T^{\mu\nu}$, which has the same appearance as the free field
energy-momentum density operator. In its calculation, only two-point Green
function is involved. What remarkable here is that the quadratic nature in $g$
of the leading term makes the vacuum energy more dependent on
the form of field interactions.

Such a hypothesis has another appealing feature. In a quantum field theory without
gravitation, a totally free field and its quanta cannot be observed
in principle because all experiments are based on interactions. So we can extend the
physical Hilbert space by taking the direct product with any free particle Hilbert space
without causing any observable effects. Employing the above hypothesis leads
to the conclusion that the vacuum of free fields does not contribute to gravitational interactions,
and it follows that direct product extension of physical Hilbert space with free particle
Hilbert spaces also does not cause any observable gravitational effects as long as
we assume no quanta are excited in such spaces considering that there are no
interactions to cause such excitations.

The hypothesis excludes the reality of zero point energy. Criticism 
may be raised based on the Casimir effect, which has been measured to
about 1\% precision~\cite{Mohideen:1998iz}. There are different opinions
on taking the Casimir effect as the evidence of zero point energy.
In ref.~\cite{Jaffe:2005vp} R.~L.~Jaffe argued that 
we can calculate Casimir forces as a dynamical effect of QED, instead of
by introducing zero point energy, and it vanishes as $\alpha$ approaches
zero. And the conclusion is drawn out that the reality of zero point energy
is not demonstrated by any known phenomenon, including the Casimir effect.
While V.~V.~Nesterenko etc.~\cite{Nesterenko:2005xv} think the Casimir effect
should be treated as the manifestation of the reality of zero point energy because
of the corresponding simpler and clearer theoretical analysis.
The criticism does not stand up if we admit the first opinion. If we accept the second
opinion, according to the theoretical calculation of Casimir effect~\cite{Itzykson},
we can still argue that what Casimir effect manifests is the zero point energy
with boundaries, i.e.~in experiments that measure Casimir effect,
the zero point energy itself is not observed, what we observe is the excess of it
caused by boundary conditions. The experimental confirmation of the Casimir effect
does not manifest the reality of the zero point energy in unbounded homogeneous space,
so it does not necessarily mean that such zero point energy has gravitational effects.

\section{The vacuum energy of QED}

Now let's turn to a real theory --- quantum electrodynamics(QED)
with  Lagrangian density
\begin{eqnarray}
  \label{eq:QED-Lagrangian}
  \mathscr{L} & = &-\frac{1}{4}F_{B\mu\nu}F_{B}^{\mu\nu}
  +\bar{\Psi}_{B} \left( i \gamma^{\mu} \partial_{\mu} -m_{B} \right)\Psi_{B}
  \nonumber \\
  & & -e_{B} \bar{\Psi}_{B}\gamma^{\mu}\Psi_{B} A_{B\mu} 
\end{eqnarray}
where the subscript $B$ indicates bare or unrenormalized quantities. In the
following, symbols without subscript $B$ implicitly denote quantities
renormalized on mass shell.
As discussed above, to first order, we only need the free energy-momentum
tensor
\begin{equation}
  T_{0}^{\mu\nu}  =  T_{\gamma,0}^{\mu\nu}+T_{e,0}^{\mu\nu}
\end{equation}
where
\begin{eqnarray}
  \label{eq:QED-emt1}
  T_{\gamma,0}^{\mu\nu} & = & F^{\sigma\mu}F_{ \phantom{\nu}\sigma }^{\nu}
  -g^{\mu\nu} \mathscr{L}_{\gamma,0} \nonumber \\
  & = & F^{\sigma\mu}F_{ \phantom{\nu}\sigma }^{\nu}
  + \frac{1}{4}g^{\mu\nu}F_{\rho\sigma}F^{\rho\sigma} \\
  \label{eq:QED-emt2}
  T_{e,0}^{\mu\nu} & = & \bar{\Psi} i \gamma^{\mu} \partial^{\nu} \Psi
  -g^{\mu\nu}\mathscr{L}_{e,0} \nonumber \\
  & = &  \bar{\Psi} i \gamma^{\mu} \partial^{\nu} \Psi 
  -g^{\mu\nu} \bar{\Psi} \left( i \gamma^{\sigma}\partial_{\sigma}-m \right)\Psi
\end{eqnarray}
Using equations like
Eq.~(\ref{eq:phi4-quadratic})~(\ref{eq:phi4-partial})~(\ref{eq:phi4-quartic})
and our hypothesis, we can begin to estimate 
the vacuum energy of QED. The estimation is done in Feynman
gauge, and considering an infrared cutoff(for example, caused by the
cosmological horizon, see~\cite{Padmanabhan:2004qc} for related discussion)
we introduce the effective photon mass $\eta$ in its propagator.

To first order~\cite{Itzykson}
\begin{eqnarray}
  & & \langle 0 \vert T \{A^{\mu}(x)A^{\nu}(y)\} \vert 0 \rangle_{l}
  \nonumber \\
  &=& \frac{i}{ \left( 2\pi \right)^{4} } \int \mathrm{d}^{4}p
  e^{-ip \cdot \left( x-y \right) } \nonumber \\
  & & \times \left(  \frac{1}{p^{2}-\eta^{2}+i\epsilon } \right)^{2}
  \left( p^{\mu}p^{\nu}-p^{2}g^{\mu\nu} \right) \pi(p^{2}) 
\end{eqnarray}
(the subscript $l$ indicates that the Green function only includes loop
graphs and corresponding counter terms) where
\begin{eqnarray}
  \pi(p^2) & = & -\frac{e^2}{12\pi^{2}} p^2
  \int_{4m^{2}}^{+\infty} \frac{\mathrm{d}k^{2}}{k^{2}}
  \frac{1}{k^{2}-p^{2}-i\epsilon} \nonumber \\
  & & \times
  \left( 1- \frac{4m^{2}}{k^{2}} \right)^{\frac{1}{2}}
  \left( 1+ \frac{2m^{2}}{k^{2}} \right)
\end{eqnarray}
Together with Eq.~(\ref{eq:QED-emt1}), we can derive the energy-momentum
tensor of the vacuum of electromagnetic field
\begin{eqnarray}
  \label{eq:QED-v-emt1}
  T_{\gamma,vac}^{\mu\nu} &=& \lim_{x-y \to 0} \frac{i}{ \left(2\pi\right)^{4} }
  \int \mathrm{d}^{4}p f(|\vec{p}|,|p^{0}|) e^{-ip\cdot \left(x-y\right) }
  \nonumber \\
  && \times \left( \frac{1}{p^{2}-\eta^{2}+i\epsilon} \right)^{2} \pi(p^{2})
  \nonumber \\
  && \times p^{2} \left( 2p^{\mu}p^{\nu}-\frac{1}{2}g^{\mu\nu}p^{2} \right)
\end{eqnarray}
where $f(|\vec{p}|,|p^{0}|)$ is a cutoff function introduced by, say,
supersymmetry and/or any other unknown extremely high energy physics.
It is expected to approximately equal $1$ if $|\vec{p}|$ and $|p^{0}|$
are small and $0$ if $|\vec{p}|$ or $|p^{0}|$ are large. For
simplicity we take
\begin{equation}
  \label{eq:QED-cutoff1}
  f(|\vec{p}|,|p^{0}|)= \left\{
      \begin{array}{ll}
        1 & \textrm{if $|\vec{p}| \le \Lambda $ and $|p^{0}| \le \Lambda$ } \\
        0 & \textrm{if $|\vec{p}| > \Lambda $ or $|p^{0}| > \Lambda$ }
      \end{array}
    \right.
\end{equation}
and assume it has no poles in the $p^{0}$ complex plane.
For the vacuum energy, i.e. $\mu=\nu=0$,
carrying out the integral over $p^{0}$, we get
\begin{eqnarray}
  T_{\gamma,vac}^{00} &=& \frac{e^{2}}{12\pi^{2}}
  \frac{1}{\left(2\pi\right)^{3}} \int \mathrm{d}^{3}p
  \int_{4m^{2}}^{+\infty}\mathrm{d}k^{2}
  f(|\vec{p}|,\sqrt{\vec{p}^{2}+k^{2}})
  \nonumber \\
  && \times
  \frac{\vec{p}^{2}+\frac{3}{4}k^{2}}{\sqrt{\vec{p}^{2}+k^{2}}}
  \frac{k^{2}}{\left( k^{2}-\eta^{2}+i\epsilon \right)^{2}}
  \nonumber \\
  && \times
  \left(1-\frac{4m^{2}}{k^{2}}\right)^{\frac{1}{2}}
  \left(1+\frac{2m^{2}}{k^{2}}\right)
\end{eqnarray}
Substitute Eq.~(\ref{eq:QED-cutoff1}) into it
\begin{eqnarray}
  \label{eq:QED-ve1}
  T_{\gamma,vac}^{00} &=& \frac{e^{2}}{3\pi}
  \frac{1}{\left(2\pi\right)^{3}}
  \int_{4m^{2}}^{\Lambda^{2}}\mathrm{d}k^{2}
  \int_{0}^{\sqrt{\Lambda^{2}-k^{2}}} \mathrm{d}|\vec{p}|
  \nonumber \\
  && \times
  \frac{|\vec{p}|^{2}\left(|\vec{p}|^{2}+\frac{3}{4}k^{2}\right)}
  {\sqrt{|\vec{p}|^{2}+k^{2}}}
  \frac{k^{2}}{\left( k^{2}-\eta^{2}+i\epsilon \right)^{2}}
  \nonumber \\
  && \times
  \left(1-\frac{4m^{2}}{k^{2}}\right)^{\frac{1}{2}}
  \left(1+\frac{2m^{2}}{k^{2}}\right)
  \nonumber \\
  &=& \frac{e^{2}}{12\pi}
  \frac{1}{\left(2\pi\right)^{3}}
  \int_{4m^{2}}^{\Lambda^{2}}\mathrm{d}k^{2}
  \Lambda \left( \Lambda^{2}-k^{2} \right)^{\frac{3}{2}}
  \nonumber \\
  && \times
  \frac{k^{2}}{\left( k^{2}-\eta^{2}+i\epsilon \right)^{2}}
  \nonumber \\
  && \times
  \left(1-\frac{4m^{2}}{k^{2}}\right)^{\frac{1}{2}}
  \left(1+\frac{2m^{2}}{k^{2}}\right)
\end{eqnarray}

Similarly, for the electron field, to first order~\cite{Itzykson}
\begin{eqnarray}
  && \langle 0 \vert T \{ \Psi(x)\bar{\Psi}(y) \} \vert 0 \rangle_{l}
  \nonumber \\
  &=& \frac{i}{\left(2\pi\right)^{4}}\int \mathrm{d}^{4}p
  e^{-ip \cdot \left(x-y\right) }
  \frac{\left(\not\!p+m\right)\Sigma^{*}(p)\left(\not\!p+m\right)}
  { \left( p^{2}-m^{2}+i\epsilon \right)^{2} }
\end{eqnarray}
and
\begin{eqnarray}
  \label{eq:electron-self-energy}
  \Sigma^{*}(p)
  &=& -\frac{e^{2}}{16\pi^{2}} 
  \Bigg\{
  \not\!p 
  \left[
    \left( 
      \frac{m^{4}}{\left(p^{2}\right)^{2}}-1
    \right)
    \mathrm{L}(p^{2})
    +\frac{m^{2}}{p^{2}}-1
  \right]
  \nonumber \\
  &&
  +4m\left(1-\frac{m^{2}}{p^{2}}\right)\mathrm{L}(p^{2})
  \nonumber \\
  &&
  -2\left(\not\!p-m\right)
  \left(1+\ln\frac{\eta^{2}}{m^{2}}\right)
  \Bigg\}
\end{eqnarray}
where
\begin{eqnarray}
  \mathrm{L}(p^{2}) &=& \left\{
    \begin{array}{ll}
      \ln\left(1-\frac{p^{2}}{m^{2}}\right) & \textrm{if $p^{2} \le m^{2}$}\\
      \ln\left(\frac{p^{2}}{m^{2}}-1\right)-i\pi & \textrm{if $p^{2} > m^{2}$}
    \end{array}
  \right. \nonumber \\
  &=& \int_{m^{2}}^{+\infty}\mathrm{d}k^{2}
  \left(
    \frac{1}{k^{2}}-\frac{1}{k^{2}-p^{2}-i\epsilon}
  \right)
\end{eqnarray}
Since
\begin{eqnarray}
  T_{e,0}^{\mu\nu} & = & \lim_{x-y \to 0} \mathrm{Tr} \bigg[ 
  -i \gamma^{\mu}\frac{\partial}{\partial x_{\nu}} 
  T \{\Psi(x)\bar{\Psi}(y)\}  \nonumber \\
  && + g^{\mu\nu}
  \left(i\gamma^{\sigma}\frac{\partial}{\partial x^{\sigma}}-m\right)
  T \{\Psi(x)\bar{\Psi}(y)\}
  \bigg]
\end{eqnarray}
according to our hypothesis, we have
\begin{eqnarray}
  T_{e,vac}^{\mu\nu} & = & \lim_{x-y \to 0} \mathrm{Tr} \bigg[ 
  -i \gamma^{\mu}\frac{\partial}{\partial x_{\nu}} 
  \langle 0 \vert T 
  \{\Psi(x)\bar{\Psi}(y)\} 
  \vert 0 \rangle_{l} \nonumber \\
  && + g^{\mu\nu}
  \left(i\gamma^{\sigma}\frac{\partial}{\partial x^{\sigma}}-m\right)
  \langle 0 \vert
  T \{\Psi(x)\bar{\Psi}(y)\}
  \vert 0 \rangle_{l} 
  \bigg] \nonumber \\
  &=& \lim_{x-y \to 0}\mathrm{Tr}
  \Bigg\{
  \frac{i}{\left(2\pi\right)^{4}}
  \int \mathrm{d}^{4}p g(|\vec{p}|,|p^{0}|) e^{-ip \cdot \left(x-y\right)}
  \nonumber \\
  && \times
  \frac{ -\gamma^{\mu}p^{\nu}+g^{\mu\nu}\left(\not\!p-m\right) }
  { \left(p^{2}-m^{2}+i\epsilon\right)^{2} } \nonumber \\
  && \times
  \left(\not\!p+m\right) \Sigma^{*}(p) \left(\not\!p+m\right)
  \Bigg\}
\end{eqnarray}
where $g(|\vec{p}|,|p^{0}|)$ is also a cutoff function like
$f(|\vec{p}|,|p^{0}|)$. Again we take a simple form of it
\begin{equation}
  \label{eq:QED-cutoff2}
  g(|\vec{p}|,|p^{0}|) = \left\{
    \begin{array}{ll}
      1 & \textrm{if $|\vec{p}| \le \Lambda$
        and $|p^{0}| \le \sqrt{\Lambda^{2}+m^{2}}$ } \\
      0 & \textrm{if $|\vec{p}| > \Lambda$
        or $|p^{0}| > \sqrt{\Lambda^{2}+m^{2}}$ }
    \end{array}
  \right.
\end{equation}
and assume it has no poles in the $p^{0}$ complex plane.
After some derivation similar to the case for the electromagnetic field, we
get
\begin{eqnarray}
  \label{eq:QED-ve2}
  && T_{e,vac}^{00} \nonumber \\
  &=& -\frac{e^{2}}{16\pi^{4}}
  \int_{m^{2}}^{\Lambda^{2}+m^{2}}\mathrm{d}k^{2}
  \frac{1}{k^{2}\left( k^{2}-m^{2} \right)}
  \nonumber \\
  && \times
  \Bigg[
  \left( \frac{m^{4}}{k^{2}}-6m^{2}+k^{2} \right)
  \int_{0}^{\sqrt{\Lambda^{2}+m^{2}-k^{2}}} \mathrm{d} |\vec{p}|
  \frac{|\vec{p}|^{4}}{\sqrt{|\vec{p}|^{2}+k^{2}}}
  \nonumber \\
  &&
  -2m^{2}\left(k^{2}+m^{2}\right)
  \int_{0}^{\sqrt{\Lambda^{2}+m^{2}-k^{2}}} \mathrm{d} |\vec{p}|
  \frac{|\vec{p}|^{2}}{\sqrt{|\vec{p}|^{2}+k^{2}}}
  \nonumber \\
  &&
  +4m^{2} \int_{0}^{\Lambda}\mathrm{d} |\vec{p}|
  |\vec{p}|^{2} \sqrt{|\vec{p}|^{2}+m^{2}}
  \Bigg]
  \nonumber \\
  &&
  -\frac{e^{2}}{16\pi^{4}}
  \left(
    4 + 2\ln\frac{\eta^{2}}{m^{2}} 
    + 4\ln\frac{\Lambda^{2}+m^{2}}{\Lambda^{2}}
  \right)
  \nonumber \\
  && \times
  \int_{0}^{\Lambda}\mathrm{d}|\vec{p}|
  |\vec{p}|^{2} \sqrt{|\vec{p}|^{2}+m^{2}}
\end{eqnarray}

\begin{figure}[htbp]
  \centering
  \includegraphics[angle=-90,width=0.5\textwidth]{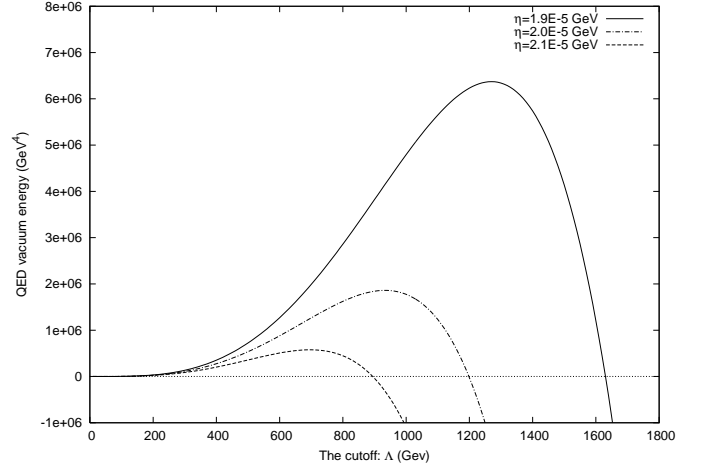}
  \caption{QED vacuum energy crosses zero at TeV energy scale}
  \label{fig:susy}
\end{figure}

\begin{figure}[htbp]
  \centering
  \includegraphics[angle=-90,width=0.5\textwidth]{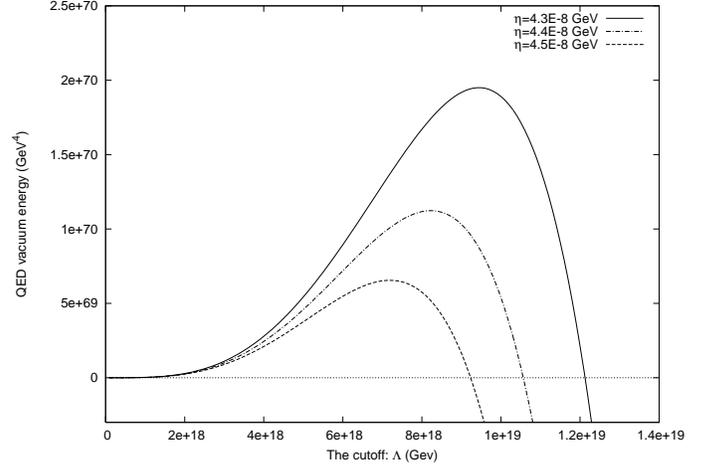}
  \caption{QED vacuum energy crosses zero at Planck energy scale}
  \label{fig:planck}
\end{figure}

Since the first order vacuum energy depends on the interaction,
we should view the vacuum energy of the electromagnetic field and that
of electron field as a whole, i.e. the first order QED vacuum
energy
\begin{equation}
  \label{eq:QED-ve}
  T_{QED,vac}^{00}=T_{\gamma,vac}^{00}+T_{e,vac}^{00}
\end{equation}
Fig.~(\ref{fig:susy}) and Fig.~(\ref{fig:planck}) show the dependence
of QED vacuum energy of different effective photon mass $\eta$ on the
cutoff $\Lambda$. We see that QED vacuum energy crosses zero at some
cutoff. This property makes it possible for us to get an arbitrarily
small vacuum energy or cosmological constant. The
cosmological constant problem could probably be resolved in this way. What we need to
do is just to show that the vacuum energy with the correct infrared
and ultraviolet cutoff is consistent with the observed cosmological
constant. Of course, the real situation, in which many other fields
should be considered, would be much more complicated.

The infrared cutoff or the effective photon mass $\eta$ most probably
arises from the cosmological horizon based on causality
consideration. Now the problem is that the length scales corresponding
to the infrared cutoffs in Fig.~(\ref{fig:susy}) and
Fig.~(\ref{fig:planck}) are too small to be comparable with the
cosmological horizon scale. There may be some mechanisms to cause such 
cutoffs, but another more reasonable possibility is that our assumptions
about the cutoff functions are too simple to be realistic
and many other fields are not included.

If the real cutoff functions derived from some theories beyond the
standard model together with a $\eta$ corresponding to the
cosmological horizon leads to a vacuum energy consistent with the
observed cosmological constant, then we can extract some useful
information, even though we do not know the explicit form
of the cutoff functions,
Investigating the dependence of the vacuum energy on
$\eta$, since for small $\eta$
\begin{equation}
  \left( \frac{1}{p^{2}-\eta^{2}+i\epsilon} \right)^{2}
  \approx
  \frac{\left(p^{2}\right)^{2}+2p^{2}\eta^{2}}{\left(p^{2}+i\epsilon\right)^{4}}
\end{equation}
we have the vacuum energy
\begin{equation}
  \label{eq:ve-vs-eta}
  \rho_{vac}(\eta)=\rho_{0}+\rho_{1}\frac{\eta^{2}}{m^{2}}
  +\rho_{2}\ln\frac{\eta^{2}}{m^{2}}
\end{equation}
or the cosmological ``constant''
\begin{equation}
  \lambda(\eta)=\lambda_{0}+\lambda_{1}\frac{\eta^{2}}{m^{2}}
  +\lambda_{2}\ln\frac{\eta^{2}}{m^{2}}
\end{equation}
We see that the cosmological ``constant'' is no more a constant, but a
parameter depends on the infrared cutoff. Taking the cosmological
horizon that blocks information as the cause of the infrared cutoff,
we will get a time-dependent cosmological ``constant'' because of the
time-dependence of the cosmological horizon.

\section{Conclusion}

From the hypothesis that under the approximation that the quantum
equations of motion reduce to the classical ones,
the quantum vacuum also reduces to the classical vacuum,
we propose that the part of vacuum energy-momentum
that arises from tree level Green functions is unobservable in principle, hence
does not contribute to gravitation. As a result, two usually accepted facts,
i.e.~the quantum corrections
to classical theory only come from loop Feynman graphs, and the direct product
extension of physical Hilbert space with free particle Hilbert spaces does not cause any
observable effects, are extended from quantum field theory to theories with gravitation
included. The hypothesis does not contradict the Casimir effect, which
has been proved experimentally.
The evaluation of the QED vacuum energy under the hypothesis shows
that it crosses zero at some cutoff, which provides a possible solution
to the cosmological constant problem. In the calculation we use
very simple cutoff functions, which should in principle be derived from theories
beyond the standard model in further studies.
From the dependence of QED vacuum
energy on the infrared cutoff, we derive a kind of
parameterization of the cosmological ``constant''.

\appendix
\section{}
For the scalar field theory, the time-ordered product is defined as
\begin{eqnarray}
  && T\left\{ \phi(x)\phi(y) \right\} \nonumber \\
  &=& \theta(x^0-y^0)\phi(x)\phi(y) + \theta(y^0-x^0)\phi(y)\phi(x)
\end{eqnarray}
We will show that
\begin{eqnarray}
  \label{eq:tobeproved}
  && \lim_{x-y \to 0}
  \frac{\partial}{\partial x^{\mu}}
  \frac{\partial}{\partial y^{\nu}}
  T\left\{ \phi(x)\phi(y) \right\}
  \nonumber \\
  &=& \lim_{x-y \to 0}
  T\left\{
    \frac{\partial}{\partial x^{\mu}}\phi(x)
    \frac{\partial}{\partial y^{\nu}}\phi(y)
  \right\}
\end{eqnarray}
If both $\mu$ and $\nu$ are space indices, $\mu=i, \nu=j$, obviously
\begin{equation}
  \label{eq:case1}
  \frac{\partial}{\partial x^{i}}
  \frac{\partial}{\partial y^{j}}
  T\left\{ \phi(x)\phi(y) \right\}
  = T\left\{
    \frac{\partial}{\partial x^{i}}\phi(x)
    \frac{\partial}{\partial y^{j}}\phi(y)
  \right\}
\end{equation}
If one of $\mu$ and $\nu$ stands for the time index, say $\mu=0, \nu=i$, then
\begin{eqnarray}
  \label{eq:case2}
  &&
  \frac{\partial}{\partial x^{0}}
  \frac{\partial}{\partial y^{i}}
  T\left\{ \phi(x)\phi(y) \right\}
  \nonumber \\
  & = &
  T\left\{
    \frac{\partial}{\partial x^{0}}\phi(x)
    \frac{\partial}{\partial y^{i}}\phi(y)
  \right\}
  \nonumber \\
  &&  +  \frac{\partial}{\partial y^{i}}
  \left\{ \delta(x^0-y^0) \left[\phi(x) \  \phi(y)\right]\right\}
  \nonumber \\
  & = &
  T\left\{
    \frac{\partial}{\partial x^{0}}\phi(x)
    \frac{\partial}{\partial y^{i}}\phi(y)
  \right\}
\end{eqnarray}
If both $\mu$ and $\nu$ are time indices, viz. $\mu=\nu=0$, we have
\begin{eqnarray}
  \label{eq:case3}
  &&
  \frac{\partial}{\partial x^{0}}
  \frac{\partial}{\partial y^{0}}
  T\left\{ \phi(x)\phi(y) \right\}
  \nonumber \\
  & = &
  T\left\{
    \frac{\partial}{\partial x^{0}}\phi(x)
    \frac{\partial}{\partial y^{0}}\phi(y)
  \right\}
  \nonumber \\
  &&  - \frac{\partial}{\partial x^{0}}
  \left\{ \delta(x^0-y^0)\left[\phi(x) \ \phi(y)\right] \right\}
  \nonumber \\
  &&  + \delta(x^0-y^0)
  \left[ \phi(x) \  \frac{\partial}{\partial y^{0}} \phi(y) \right]
  \nonumber \\
  & = & 
  T\left\{
    \frac{\partial}{\partial x^{0}}\phi(x)
    \frac{\partial}{\partial y^{0}}\phi(y)
  \right\}
  \nonumber \\
  &&  + i\delta^{(4)}(x-y)
\end{eqnarray}
(\ref{eq:tobeproved}) follows from (\ref{eq:case1}), (\ref{eq:case2}) and (\ref{eq:case3}),
so we have Eq.~(\ref{eq:phi4-partial})

Similarly, for the spinor field theory
\begin{eqnarray}
  T\{ \Psi_{\alpha}(x) \bar{\Psi}_{\beta}(y) \}
  & = & \theta(x^0-y^0)\Psi_{\alpha}(x) \bar{\Psi}_{\beta}(y)
  \nonumber \\
  && - \theta(y^0-x^0)\bar{\Psi}_{\beta}(y)\Psi_{\alpha}(x)
\end{eqnarray}
where $\alpha, \beta$ denote Dirac indices, we need to prove
\begin{eqnarray}
  \label{eq:spinor}
  && \lim_{x-y \to 0}
  \frac{\partial}{\partial x^{\mu}}
  T\left\{ \Psi_{\alpha}(x) \bar{\Psi}_{\beta}(y) \right\}
  \nonumber \\
  &=& \lim_{x-y \to 0}
  T\left\{
    \frac{\partial}{\partial x^{\mu}}
    \Psi_{\alpha}(x) \bar{\Psi}_{\beta}(y)
  \right\}
\end{eqnarray}
which is obviously true for $\mu=1,2,3$. For $\mu=0$
\begin{eqnarray}
  && \frac{\partial}{\partial x^0}
  T\left\{ \Psi_{\alpha}(x) \bar{\Psi}_{\beta}(y) \right\}
  \nonumber \\
  & = &
  T\left\{
    \frac{\partial}{\partial x^0}
    \Psi_{\alpha}(x)  \bar{\Psi}_{\beta}(y)
  \right\}
  \nonumber \\
  && + \delta(x^0-y^0)
  \left[ \Psi_{\alpha}(x) \ \bar{\Psi}_{\beta}(y) \right]_{+}
  \nonumber \\
  & = &
  T\left\{
    \frac{\partial}{\partial x^0}
    \Psi_{\alpha}(x)  \bar{\Psi}_{\beta}(y)
  \right\}
  \nonumber \\
  &  &
  + \gamma^0_{\alpha\beta} \delta^{(4)}(x-y)
\end{eqnarray}
so we have Eq.~(\ref{eq:spinor}). 

\bibliographystyle{apsrev}
\bibliography{ccp}

\end{document}